\documentclass[useAMS,usenatbib]{mn2e}
\usepackage{graphicx}
\usepackage[a4paper]{hyperref}
\usepackage{amssymb}

\newcommand\aj{AJ}%
\newcommand\araa{ARA\&A}%
\newcommand\apj{ApJ}%
\newcommand\apjl{ApJ}%
\newcommand\apjs{ApJS}%
\newcommand\apss{Ap\&SS}%
\newcommand\aap{A\&A}%
\newcommand\pasp{PASP}%
\newcommand\memsai{Mem.~Soc.~Astron.~Italiana}%
\title[Blue-IR Light Curves of V582 Mon (KH 15D) from 1955 to 1970] {Blue and IR Light Curves
of the Mysterious Pre-Main Sequence Star V582 Mon (KH 15D) from
1955 to 1970}
\author[Maffei, Ciprini, Tosti]
    {Paolo Maffei$^{1}$, Stefano Ciprini$^{1,2}$\thanks{offprints: stefano.ciprini@pg.infn.it},
     and Gino Tosti$^{1}$\\
$^1$Physics Department and Astronomical Observatory, University
of Perugia, via Pascoli, 06123 Perugia, Italy\\
  $^2$Tuorla Astronomical Observatory, University of Turku,
V\"{a}is\"{a}l\"{a}ntie 20,
    21500 Piikki\"{o}, Finland\\
  }
\date{submitted to MNRAS}

\pagerange{\pageref{firstpage}--\pageref{lastpage}} \pubyear{2004}

\begin{document}

\label{firstpage}

\maketitle

\begin{abstract}
In recent years, an increasing number of publications have been
addressed to the peculiar and mysterious pre-main sequence star
V582 Mon, also known as KH 15D. This extraordinary T Tauri star,
located in the young star cluster NGC 2264, appears as to be an
eclipsing variable. In the present paper, we report a unique and
self-consistent set of light curves in the blue and near-infrared
bands, spanning a 15-year interval (epoch 1955-1970). Our
photometric data show clearly the beginning of the eclipse stage
occurred in early 1958 in the blue, and perhaps around four years
later in the infrared. The light curve period turns out to be the
same reported by recent observations (about 48.3 days), so that no
evidence for a period change results. On the other hand, in our
data the light curve shape appears as sinusoidal and is therefore
different from the one displayed today. The photometric behaviour,
determined with time-series and colour-index analysis, suggests
that V582 Mon (KH 15D) could be initially surrounded by an
accretion disk/torus seen edge-on, with subsequent thin dust
formation at the beginning of the blue radiation absorption. The
dust could then aggregate into larger particles providing the
transition between selective and total absorption, accompanied
with eclipsing variability in the infrared. The minima of the
periodic light curve become deeper due to the increasing dimension
and number of dust grains, and then flattens due to a contraction
in the disk.
\end{abstract}
%
%
\section{Introduction}\label{sec:intro}
The variable star V582 Mon
($\alpha=6\mathrm{h}~41\mathrm{m}~10.3\mathrm{s},~
\delta=+09^{\circ}~28'~34''~(\mathrm{J2000.0}))$ was discovered by
\citet{Badalian72}. It is classified as a type INS (i.e. Orion
variable with rapid light changes: up to $1$ mag in 1--10 days)
in the GCVS \citep{Kholopov85}. This star was re--discovered as an
eclipse variable by \citet{Kearns98} during a systematic study of
young star variability in the region of the cluster
NGC 2264.
%
\begin{table}
\caption[]{The number of Blue and IR photographic plates in the
field of V582 Mon (KH 15D) from 1955 to 1970.} \label{tab:numobs}
\vspace{-0.2cm}
\centering {}
\par
\normalsize{
\begin{tabular}{l}\hskip -0.37cm
{\resizebox{8.7cm}{!}{\includegraphics{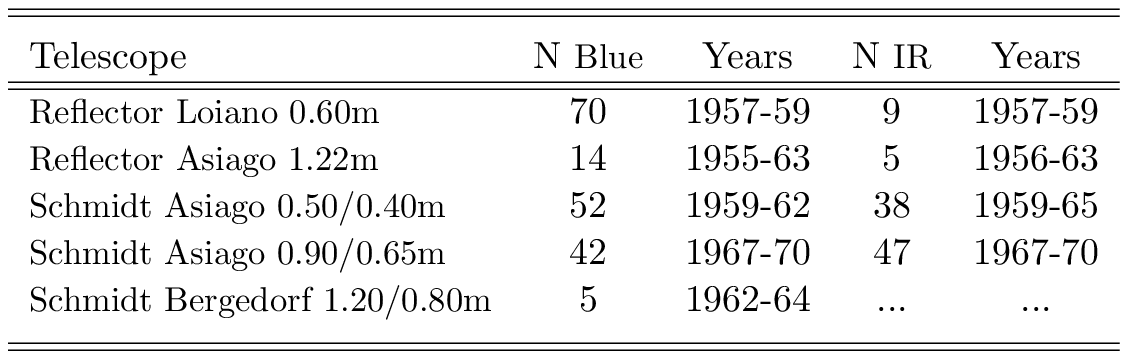}}}\\
\end{tabular}
}
\end{table}
\normalsize
%
The star, named 15D in a list of 209 objects \citep{Kearns97},
appeared as an eclipsing variable with an amplitude of about 3 mag
in the $I_c$ (Cousins) band, with a period of 48-49 days. This was
a unique case among the very young stars observed as discussed by
\citet{Kearns97}. The star is also designated as n. 150 in the
list of \citet{Park00} and as n. 391 in the list of
\citet{Flaccomio99}.
\par The depth (3.5 mag) and the long duration of the eclipse
(about 40\% of each cycle, which lasts for 48.4 days) rule out a
companion star as the cause of this event
\citep{Hamilton01,Herbst02}. Polarization levels of 2\% observed
at half eclipse and $0.2\pm0.2$ observed out of the eclipse
\citep{Agol04}, would imply that the obscuring material is
composed of rather large particles.
%
\begin{figure*}
\begin{center}
\begin{tabular}{l}\hskip -0.3cm
{\resizebox{\hsize}{!}{\includegraphics{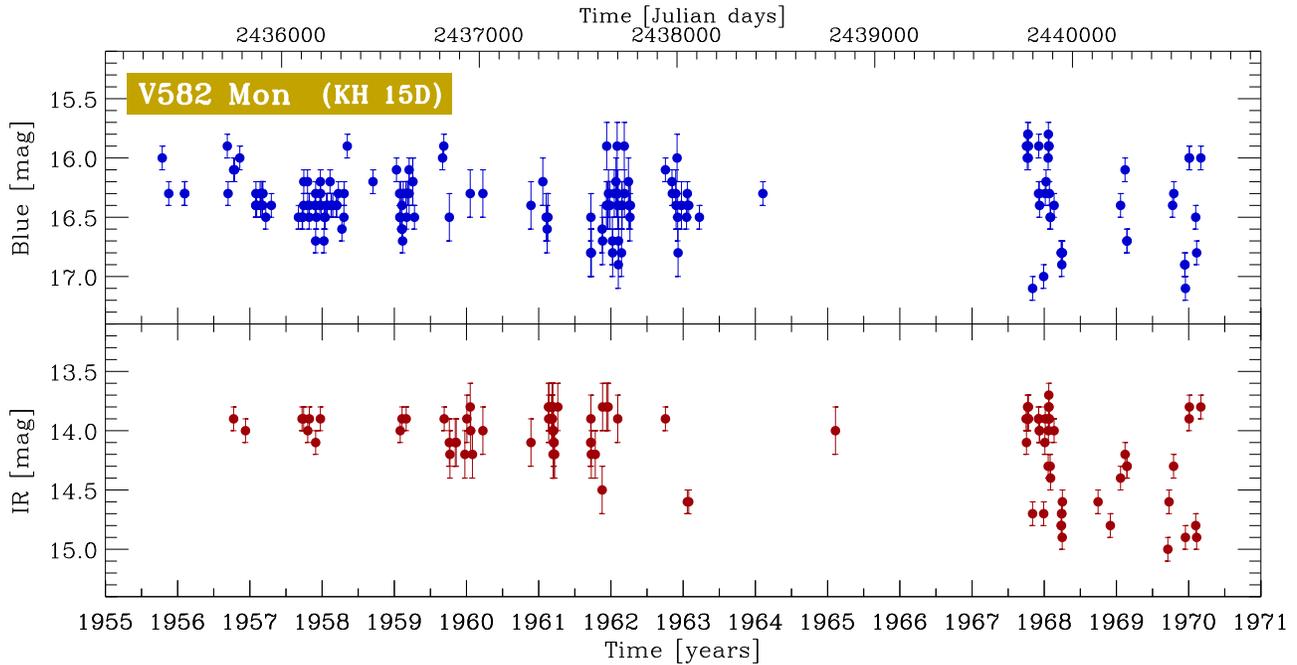}}}\\
\end{tabular}
\end{center}
\vskip -0.1 true cm \caption{The complete light curves in the blue
($\sim B$ band) and photographic IR ($\sim I_c$ band) of V582 Mon
(KH 15D) from 1955 to 1970 (due to the large plot scale
nearby observations appear as a single point).}
\label{fig:lightcurve-all}
\end{figure*}
%
In order to explain these phenomena, two hypotheses were
formulated: 1) a double star obscured in different ways by
circumstellar matter; 2) a single star surrounded by a peculiarly
warped or ridged proto-planetary disk. In the second case, the
light of the star is periodically intercepted by this dusty ridge
orbiting around the star in a plane placed edge-on with respect
to the line of sight.
\par Actually, there is a large consensus that the eclipses are caused by
circumstellar material, but it is not clear what is the
composition and the spatial distribution of this material.
%
\begin{table*}
\caption[]{Our photometric measurements in the blue and IR of V582
Mon (KH 15D) in the period 1955-1970 (for the meaning of the
telescope acronyms see Table \ref{tab:numobs}).} \label{tab:data}
\vspace{-0.2cm} \centering {}
\par
\normalsize{
\begin{tabular}{l}\hskip -0.4cm
{\resizebox{18.cm}{!}{\includegraphics{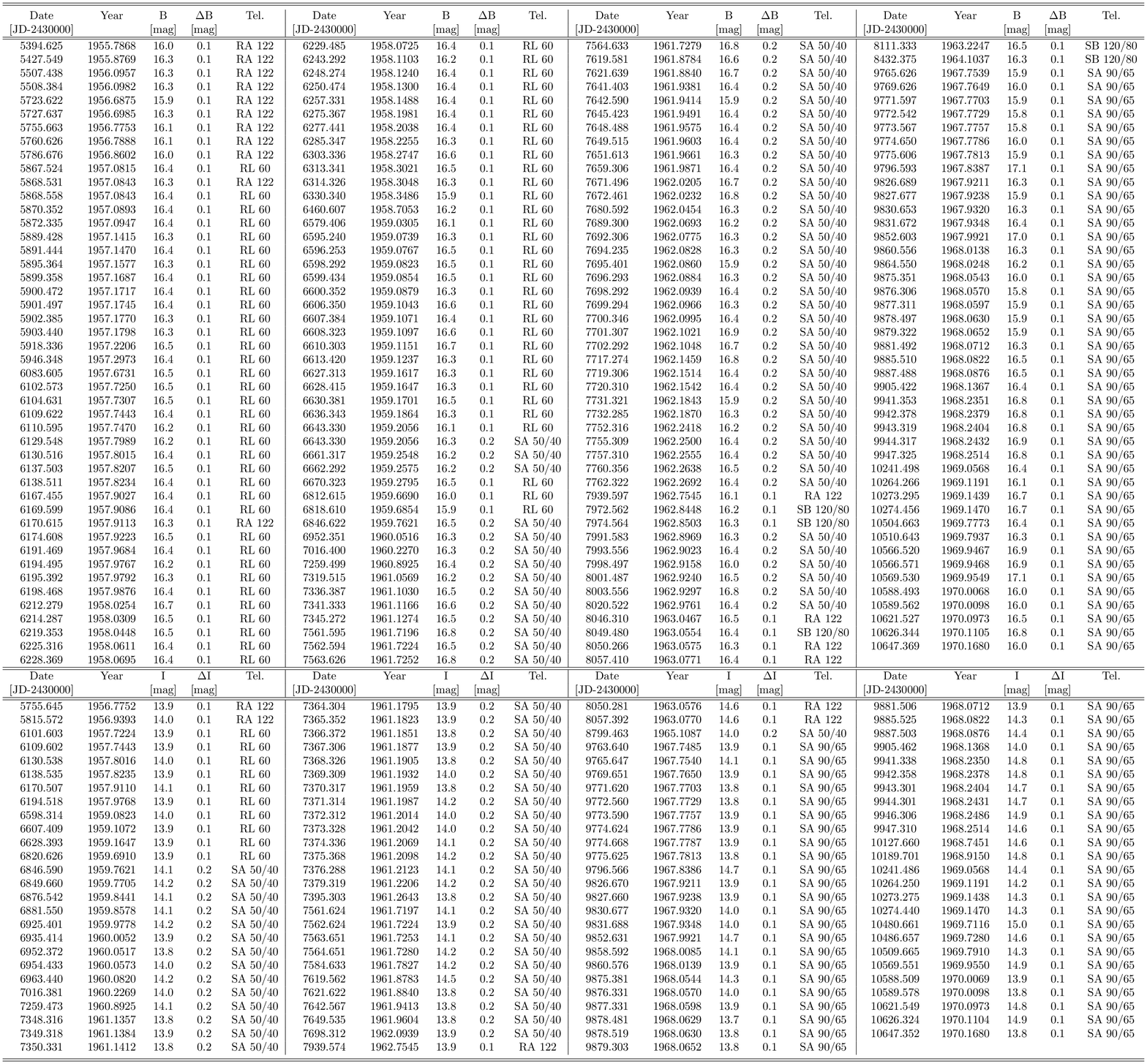}}}\\
\end{tabular}
}
\end{table*}
\normalsize
%
Theories include an edge-on circumstellar disk with a mass density
wave or a disk deformation induced by a proto-planet
\citep{Hamilton01,Herbst02,Winn03,Agol04}, an orbiting vortex of
solid particles \citep{Barge03}, and an asymmetric common envelope
\citep{Grinin02}. V582 Mon (KH 15D) is quite unique and extremely
interesting because the fortuitous alignment may allow us to study
the circumstellar (or even protoplanetary) processes, and because
the eclipse provides a natural "coronagraph" that can be used to
map out the environment of the underlying T Tauri star
\citep{Hamilton03,Agol04,Deming04}. In fact, this object should be
a weak-line T Tauri star, still accreting matter from a
circumstellar disk and endowed with a well-collimated bipolar jet
\citep{Hamilton01}. High resolution near-IR spectroscopy revealed
the presence of prominent molecular hydrogen emission probably due
to shock excitation of the ambient gas by a bipolar outflow from
the circumstellar disk \citep{Deming04}. As evidence of this
hypothesis, there is also the observation of \citet{Tokunaga04},
who revealed clearly an H$_2$ emission filament next to the
central star. Furthermore, spectro-polarimetric observations
suggest that the star is completely eclipsed, so that the flux
during the eclipse is entirely due to scattered light. The
scattering should be due to large dust grains ($\sim 10 ~\mu$m)
similar to the interplanetary grains that scatter the zodiacal
light \citep{Agol04}. Recent high-resolution spectroscopic
monitoring showed that V582 Mon is a single-line spectroscopic
binary \citep{Johnson04prep}. A model of gradual opaque-screen
occultation was proposed by \citet{Winn04}, and a model with an
eccentric binary that is encircled by a vertically thin, inclined
ring of dust and gas was suggested by \citep{Chiang04} in order to
explain the available historical light curve prior to the present
work.
\par The light curve of V582 Mon (KH 15D) from 1967 to 1982, based on
the Asiago Observatory infrared plates, displays a periodic
alternation between bright and faint states with a constant period
of 48.35 days \citep{Johnson04}. An international observation
campaign during 2001-2002 showed clearly a periodic variability
recurring with a period of $48.36$ days \citep{Herbst02}.
Observations during six seasons from 1995 to 2001 showed again a
periodical variability like an eclipsing binary, with a period of
$48.34$ days \citep{Hamilton01}. On the other hand, the analysis
of the Harvard College Observatory archival plates from 1913 to
1951 showed that there was no variability, and the deep eclipses
observed today did not occur in the first half of the 20th century
\citep{Winn03}. In that work, the need for further, more recent
archival images was underlined, as they might have shown the
onset of eclipses.
\par In our paper, we present the results of the analysis
of the plate images obtained by one of us (PM) during 15 years in
a systematic study of the variability of young stars in the
cluster NGC 2264. The data presented and discussed here pertain
to the star V582 Mon (KH 15D) and concern the observations
performed in the blue ($B$) and near-infrared ($I$) bands between
October 13, 1955 and March 1, 1970. With these data, it is
possible to determine when and how the eclipse phenomenon
occurred, its first evolution, and the subsequent dramatic
variations.
\par In Section \ref{sec:observations}, we describe the telescopes used, the photographic
plates obtained, and the data reduction. In Section
\ref{sec:dataanalysis}, we show results from the stationary period
up to 1970, and in Section \ref{sec:conclusions}, we present a
brief discussion and our conclusions.
%
\section{Observations}\label{sec:observations}
The field of the young cluster NGC 2264 was photographed in the
winter season 1955-56 with the 0.6m Loiano (Bologna, Italy) and
the 1.2m Asiago (Vicenza, Italy) telescopes. These observations
led to the discovery of 33 new variable stars, studied jointly
with the other known 78 variable stars \citep{Rosino57}. In this
systematic study, V582 Mon (KH 15D) was not detected as a
variable star.
\par In the subsequent years, the search was extended by PM to the neighboring
region of this cluster. In 15 years, 183 blue and 99 photographic
infrared plates were obtained, with five different telescopes.
Table \ref{tab:numobs} shows the distribution of the photographic
material collected. The number of plates used does not include
some bad plates (e.g. scratched, or partially lighted, or with a
covering halo due to the bright star in the field), and it does
not comprise few underexposed plates (in which the magnitude of
the variable star was much fainter than the limit-magnitude of the
plate). Hence these latter plates (about ten in $B$) were not
useful to set meaning upper limits.
\par The merit of the material used in this work lays in its
uniformity. In the blue, an emulsion 103a-O without filter was
used. In the photographic-infrared, an emulsion I-N
hypersensitized with an RG5 filter was used. The I-N/RG 5
combination transmits the spectral range 6800-8800 \AA, and has a
transmission curve similar to the 7000-9000 \AA$~$ bandpass of the
Cousins $I_c$ band. Moreover, a large fraction of the plates
(about 60\%, see Tab. \ref{tab:numobs} and Tab. \ref{tab:data})
were obtained with the 0.60m Loiano reflector and the 0.90/0.65m
Asiago Schmidt telescope, having the same focal length (210 cm)
and almost the same focal ratio (f/3.50 and f/3.23 respectively).
\par The irregular variability of the studied stars
required a high frequency in the plate collection. As a
consequence, for V582 Mon (KH 15D), a series of long duration
observations (except for the seasonal, weather and full-moon
interruptions) are available, in the years when the star began to
show its variability and during the first evolution phases. For
the first years this is true especially for the blue plates; for
the subsequent years also in the infrared ones.
\par The magnitudes of comparison stars were taken
from the blue photoelectric and photographic sequences of
\citet{Walker56}, after removing stars that appeared variable
afterwards. The calibrations curves were constructed using the
best two plates among the five ones obtained with the Schmidt
telescope of the Hamburg-Bergedorf Observatory. The infrared
comparison sequence was assembled using stars having a know colour
index $B-V$ \citep{Walker56} by means of the transformation
formulae given by \citet{Allen55}, taking care to choose only the
non-reddened stars from the colour-colour-diagram $U-B$ vs $B-V$.
\par The magnitudes of V582 Mon (KH-15D) were determined through visual
estimates with respect to its closest comparison stars. A
stereoscopic Zeiss microscope at low and variable magnification
was used. This method allows a comparison of the measurememts
obtained with different telescopes. For example, in the plates
obtained with the Loiano reflector, the star falls in a region
affected by coma, however the eye can perform a good estimation
when comparison stars of comparable coma are chosen. On the basis
of the thousands of magnitude estimations done with the above
mentioned instruments, we estimate the mean errors in the
magnitude estimations are: $\pm 0.1$ mag for the 0.6m Loiano
reflector plates, $\pm 0.05/0.1$ mag for the 1.2m Asiago reflector
plates, $\pm 0.2$ and $\pm 0.1$ mag for the two 0.50/0.40m and
0.90/0.65m Asiago Schmidt telescope plates respectively.
%
%
\section{Data Analysis and Results}\label{sec:dataanalysis}
%
Part of the analysis of the plates taken around NGC 2264 was
published in \citet{Maffei66a,Maffei66b}. The detailed study of
the more than 200 variable stars identified in the area of NGC
2264 (about one square degree) is an ongoing project. At present,
this effort has been devoted to the variable V582 Mon (KH-15D).
%
\begin{figure}
\begin{center}
\begin{tabular}{l}\hskip -0.7cm
{\resizebox{9.2cm}{!}{\includegraphics{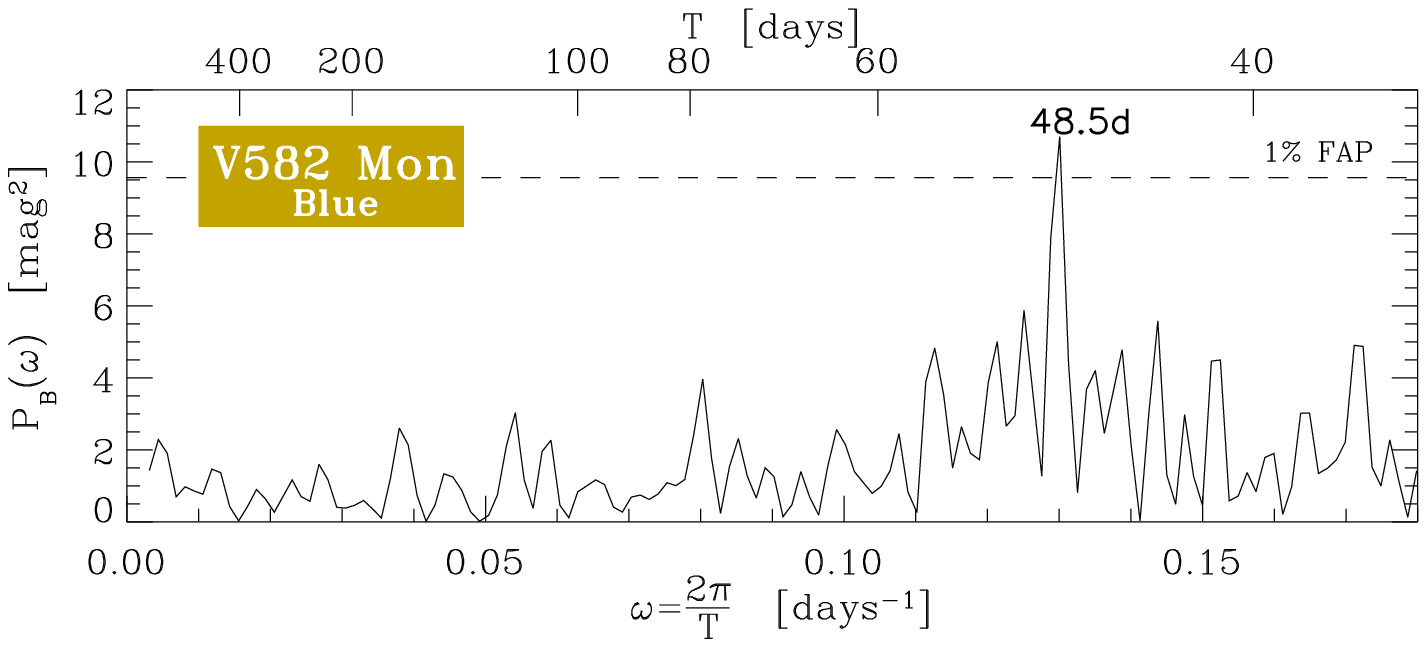}}}\\
\hskip -0.5cm
{\resizebox{9.cm}{!}{\includegraphics{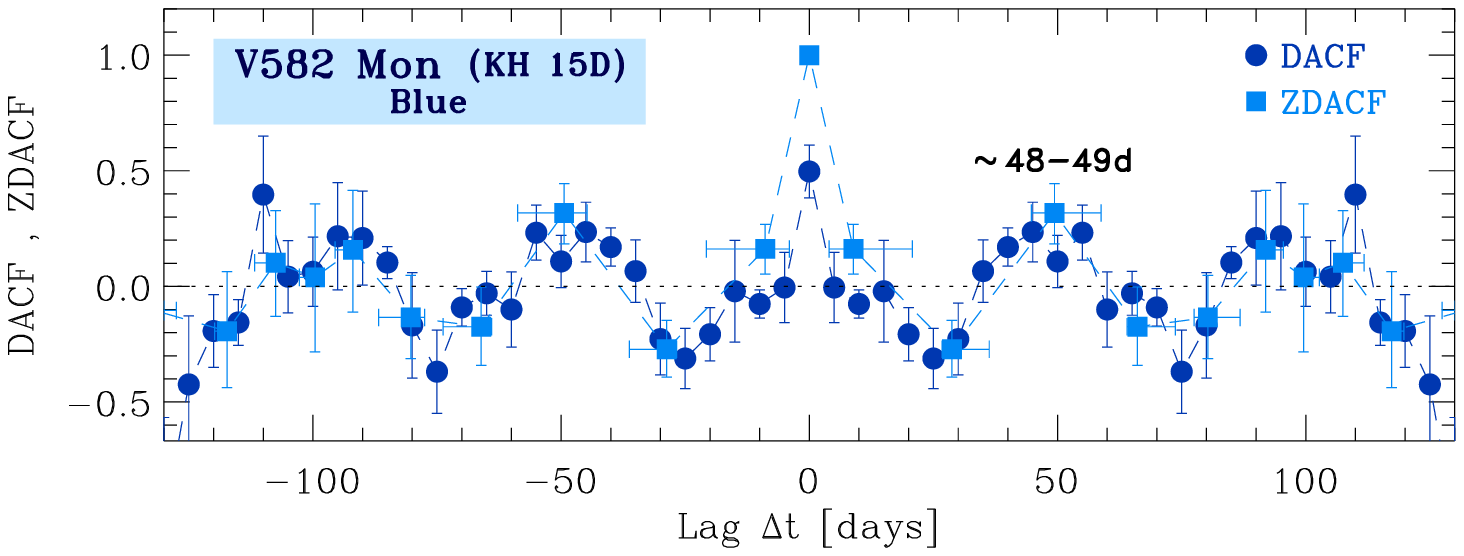}}}\\
\hskip -0.5cm
{\resizebox{9.cm}{!}{\includegraphics{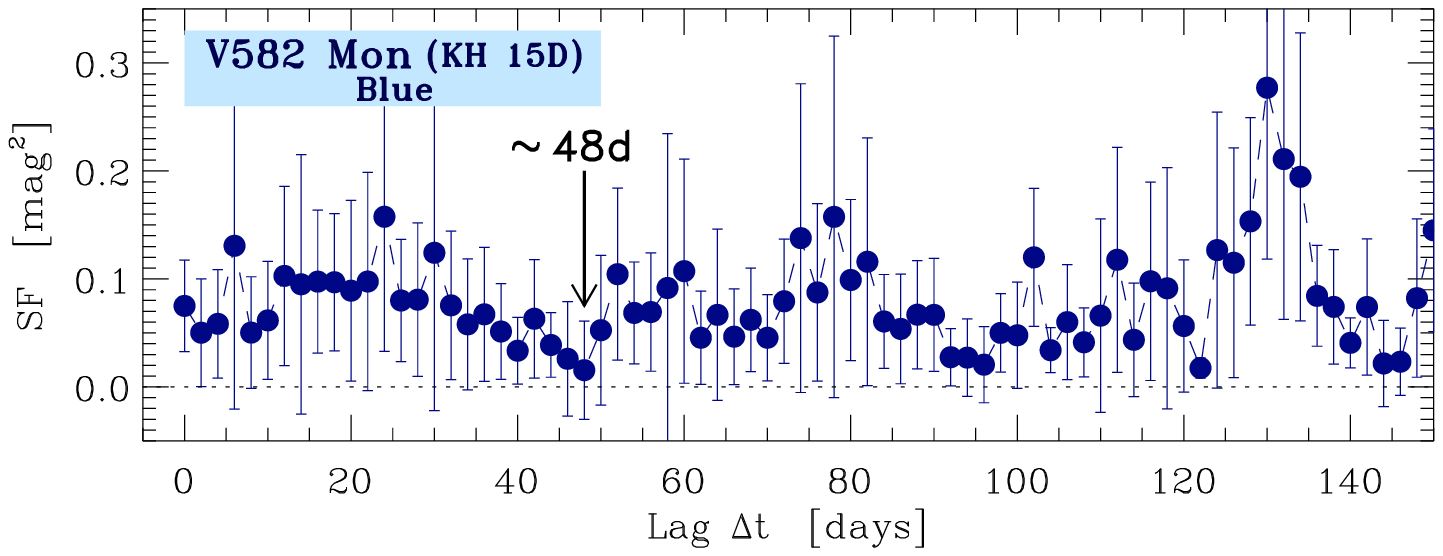}}}\\
\hskip -0.5cm
{\resizebox{9.cm}{!}{\includegraphics{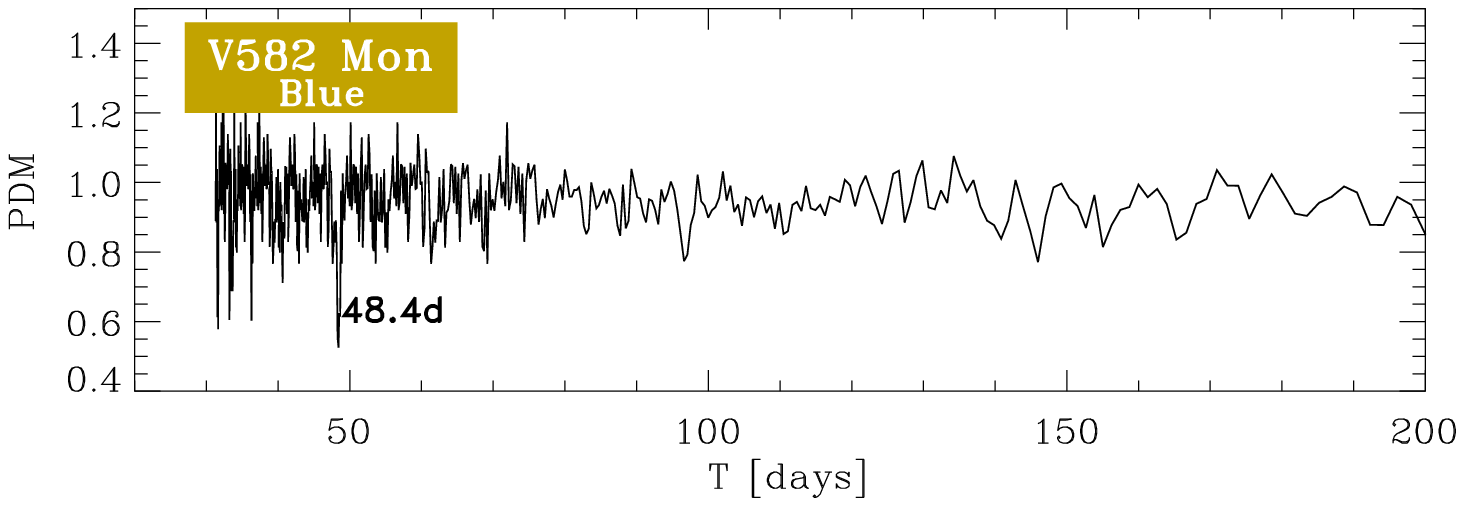}}}\\
\end{tabular}
\end{center}
\vskip -0.3 true cm \caption{Uppermost panel: periodogram plot of
the blue light curve in the best sampled interval JD 2436102
(1957.725) to JD 2438111 (1963.225). Dotted line shows the 1$\%$
false alarm significance level, under the hypothesis of
fluctuations dominated by Poisson statistics. The timescale
corresponding to the significative peak is $48.5\pm0.3$ days, in
agreement with the phase-plots findings. Second panel:
auto--correlations of the blue light curves, calculated with the
DACF and ZDACF. These show a weak broad peak at around 48-49 days.
Third panel: the first order structure function (SF) of the data
in the same period, the SF indicates a drop around a timescale of
48 days. Bottom panel: the phase dispersion minimization (PDM)
curve of the same data. The significant drop in the PDM gives a
period of $48.4\pm0.2$ days, in agreement with the previous
results. The results obtained with the different statistical
methods indicate a period equal to the actual value.}
\label{fig:periodogram_dacf}
\end{figure}
%
\begin{figure}
\begin{center}
\begin{tabular}{l}\hskip -0.7cm
{\resizebox{9.2cm}{!}{\includegraphics{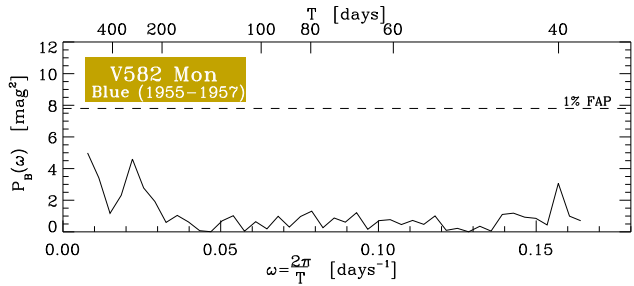}}}\\
\end{tabular}
\end{center}
\vskip -0.3 true cm \caption{Periodogram plot of the blue light
curve in the interval JD 2435394 (1955.787) to JD 2436198
(1957.988). Dotted line shows the 1$\%$ false alarm significance
level, under the hypothesis of fluctuations dominated by Poisson
statistics. The curve does not display any characteristic
timescale or periodical feature. Taking into account the good data
sampling in 1957 (see Table \ref{tab:data}) and the results
displayed in Fig. \ref{fig:periodogram_dacf}, this implies that
the variability/periodicity in the blue began in the first days of
1958.
} \label{fig:periodogram_blue19551957}
\end{figure}
%
%
\begin{figure}
\begin{center}
\begin{tabular}{l}\hskip -0.7cm
{\resizebox{9.2cm}{!}{\includegraphics{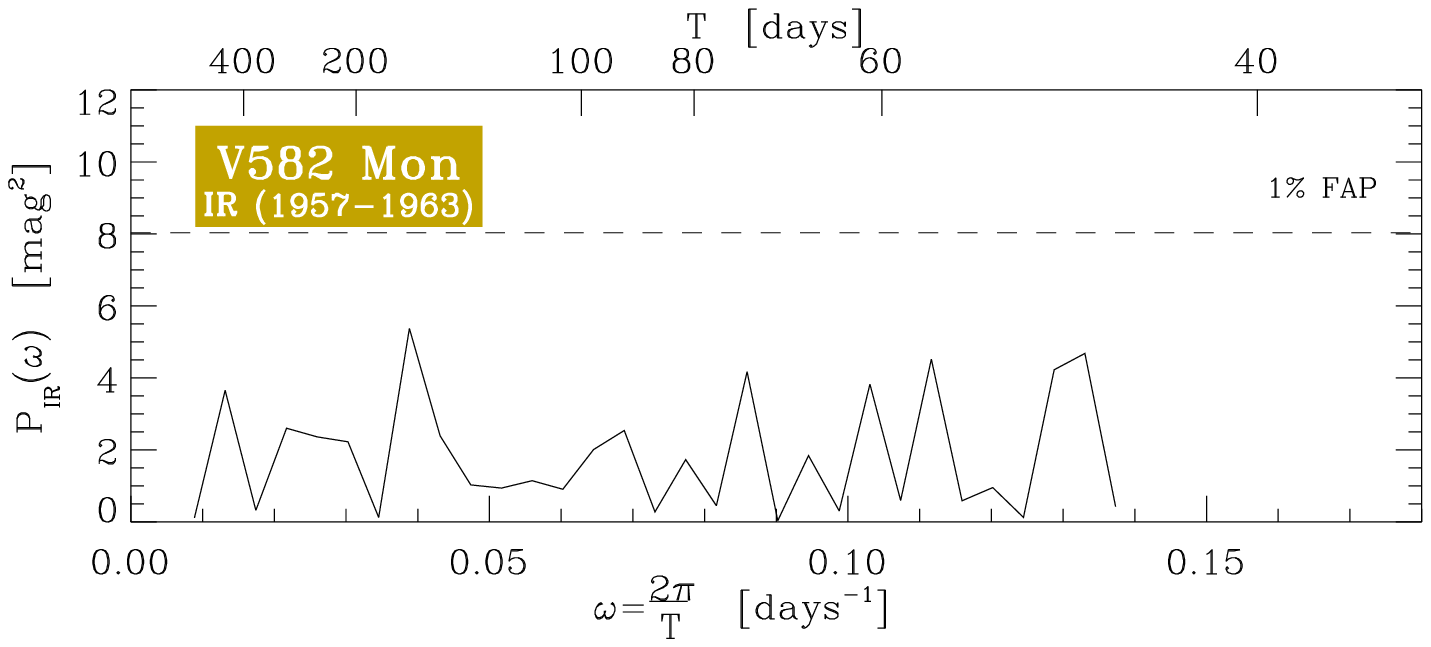}}}\\
\hskip -0.5cm
{\resizebox{9.cm}{!}{\includegraphics{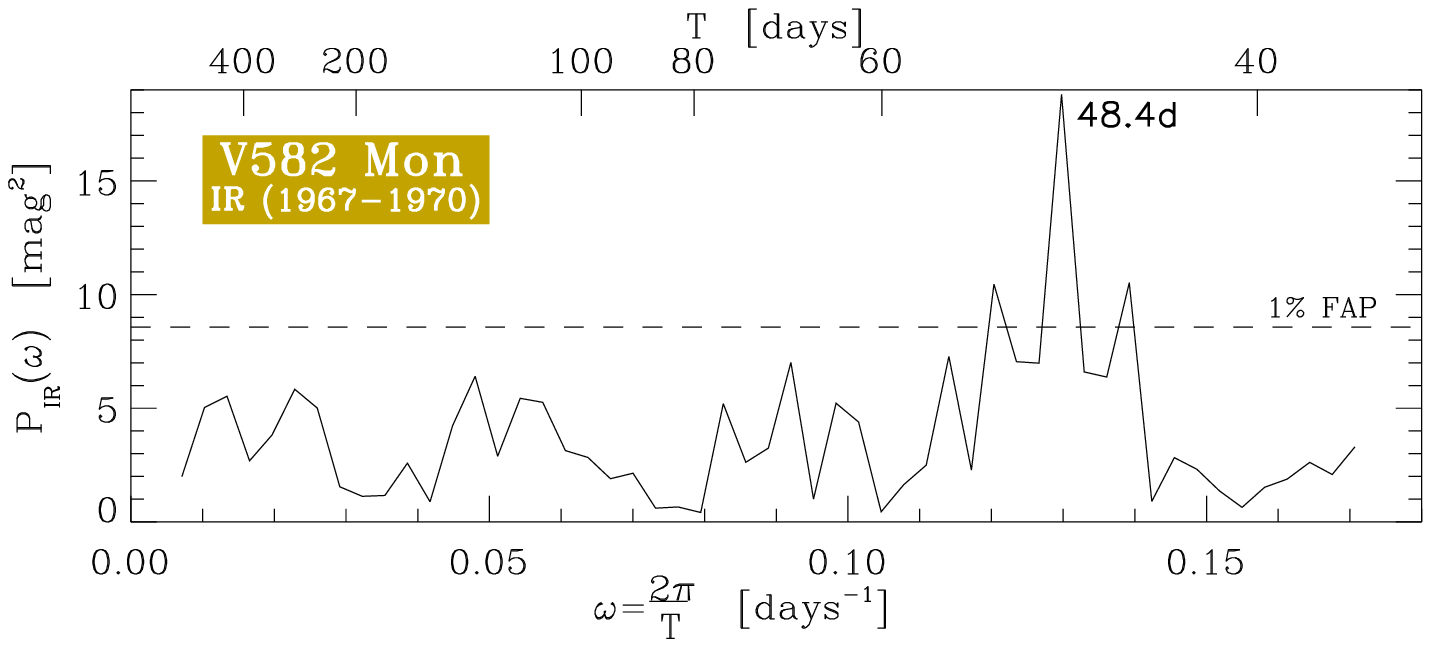}}}\\
\end{tabular}
\end{center}
\vskip -0.3 true cm \caption{Upper panel: periodogram plot of the
infrared light curve in the interval JD 2436101 (1957.722) to JD
2438057 (1963.077). Dotted line shows the 1$\%$ false alarm
significance level, under the hypothesis of fluctuations dominated
by Poisson statistics. No comparable indication of a periodicity
is seen in the infrared light curve periodogram. Lower panel:
periodogram plot of the infrared light curve in the period
1967-1970. In this interval, the characteristic periodicity (48.4
day) is clearly observed also in the infrared data.}
\label{fig:periodogram_IR}
\end{figure}
%
The infrared plates obtained with the two Asiago Schmidt
telescopes were made available to J. A. Johnson, who analyzed the
data and published a first set of results in \citet{Johnson04}.
\par As shown in Tables \ref{tab:numobs} and  \ref{tab:data}, the blue light
curve is better sampled. However for the years 1967-70, the number
of useful plates is nearly equal for the two colours, since after
the positive results obtained with the first observations in the
infrared, PM decided to take both blue and IR images for every
available observing night.
\par In the blue, V582 Mon appears with a constant
magnitude from 1955 to 1957 (Fig. \ref{fig:lightcurve-all}). The
first clear sign of a significant weakening (0.4 mag) was observed
on January 8, 1958. In the infrared, the star appeared with
constant magnitude in the subsequent 4 years, and a first sure
weakening (0.5 mag) is observed on November 16, 1961 (Fig.
\ref{fig:lightcurve-all}).
%
%
\subsection{Temporal Analysis}\label{sec:subsec:tempanalysis}
In addition to the classical phase analysis, the temporal
analysis of the data was carried out using well tested and robust
techniques suitable for an unevenly sampled time series: the
Lomb-Scargle periodogram (LSP), the first order structure function
(SF), the discrete auto-correlation function (DACF), and the phase
dispersion minimization (PDM) method . These methods allow to
investigate quantitatively the statistical temporal structure of
variability, the auto-correlations, and the existence of
characteristic time scales and periodicity.
\par The SF is equivalent to a power spectrum calculated in the time domain. It
provides information on the type of variability and the range
of the characteristic time scales that contribute to the
fluctuations \citep[see, e.g. ][]{Rutman78,Simonetti85}. Steep
drops in the SF indicate small variance and provide the signature of
possible characteristic time scales, while the slope in the
log-log representation gives the value of the power law spectrum
index, and thus the nature of the variability. The DACF allows to
investigate the level of auto-correlation in unevenly sampled data
sets \citep[see, e.g. ][]{Edelson88,Hufnagel92} without any
interpolation. The number of points per time bin can vary greatly
in the DACF, but data bins by equal population can be built
together with Montecarlo estimations for peaks and uncertainties
as in the Fisher $z$-transformed DACF
\citep[ZDACF,][]{Alexander97}. The LSP is a technique analogous to
the Fourier analysis for discrete unevenly sampled data trains,
useful to detect the strength of harmonic components with a
certain angular frequency $\omega=2\pi f$ \citep[see, e.g.
][]{Scargle82,Horne86}. Finally, the PDM method \citep[see, e.g.
][]{Stellingwerf78,Jurkevich71} seeks to minimize the variance of
the data at constant phase with respect to the mean value of the
light curve. The PDM method has no preference for a particular
periodical shape, it incorporates all the data directly into the
test statistic and it is thus well suited to small and randomly
spaced samples. A value is statistically significant when the PDM drops
near zero.
\par In Fig. \ref{fig:periodogram_dacf} we plot the LSP,
DACF and ZDACF, SF, and PDM for the blue light curve
data in the best sampled interval JD 2436102 (1957.725) to JD
2438111 (1963.225). The timescale corresponding to the
significative peak is $48.5\pm0.3$ days for the LSP. The DACF
and ZDACF also show a weak (value $\sim 0.3$) peak for a time lag
at around 48-49 days.
Both the SF and the PDM show a drop around the same value.
\par The apparently stable periodicity of the fluctuation is first
observed in 1958. The period is in good agreement with that found
in recent years with well sampled data. On the other hand the blue
light curve does not display any characteristic timescale or
periodical feature before 1958 (Fig.
\ref{fig:periodogram_blue19551957}), reinforcing the evidence for
a beginning of the eclipses around the first months of 1958.
\par The temporal analysis of the infrared light curve for
1957-1963 does not show any significant feature above statistical
fluctuations in the LSP, DACF, SF, and PDM plots (upper panel of
Fig. \ref{fig:periodogram_IR}), which indicates no corresponding
variability and periodicity in the infrared in these years. We
believe that the variability occurred only some years later
(possibly about 4 years later) in the infrared based on the
periodical behaviour in the IR displayed in the last interval of
our light curve (lower panel of Fig. \ref{fig:periodogram_IR}),
and by the $B-I$ colour analysis in Par. \ref{sec:colorindex}.
%
\subsection{Blue Light Curve}\label{sec:subsec:blueLC}
After the first weakening, the variability amplitude increased
until about 1.2 mag from the end of 1967 to the beginning of 1970.
Applying the phase periodical analysis, the maximum probability
period is 48.5 days, very similar to the ephemeris
outlined by \citet{Herbst02}. The phase light curve constructed
with the observations from 1967 to 1970, adopting the ephemeris of
\citet{Herbst02}, shows a marked regularity.
\par
\begin{figure}
\begin{center}
\begin{tabular}{l}\hskip -0.6cm
{\resizebox{9cm}{!}{\includegraphics{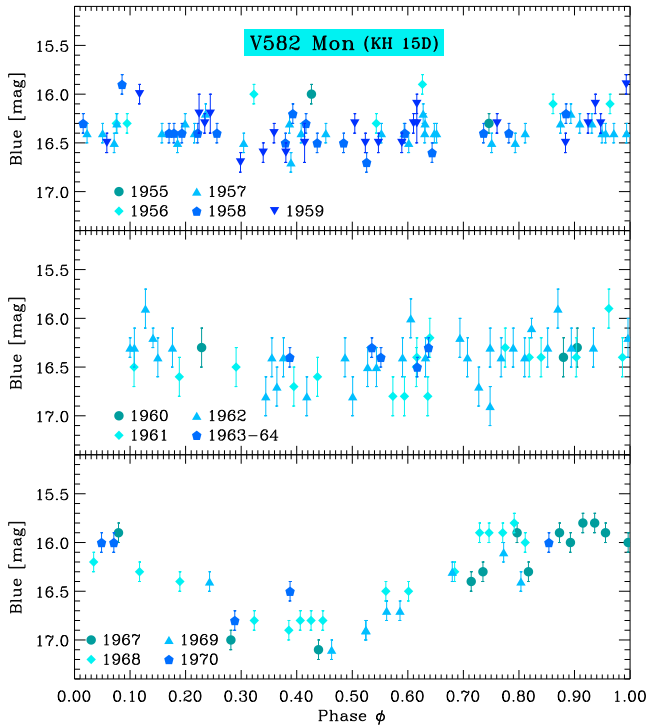}}}\\
\end{tabular}
\end{center}
\vskip -0.3 true cm \caption{Optical blue magnitude light curve of
V582 Mon (KH 15D) as function of the phase $\phi$ from our
1955-1970 data. The spectroscopic binary period (48.38 days) and
the epoch 2452352.26 reported by \citet{Herbst02} are used. All
the data points reported in the plot are real observations,
without any data point duplication. The eclipses appears to start
at the end of the 1950s. The periodic variability in the
subsequent years up to the 1970 seems rather sinusoidal, with slow
descendent and increasing stages and no plateaus.}
\label{fig:phaseplotBLU}
\end{figure}

 Figure \ref{fig:phaseplotBLU} shows the optical
blue magnitude light curve of V582 Mon (KH 15D) as a function of
the phase $\phi$ for our data. All the data points reported in the
plot are real observations. The eclipses appear clearly to start
at the end of the 1950s.
%
The periodic variability in the subsequent years up to the 1970
seems rather sinusoidal, with slow decreasing and increasing
stages and no evident plateaus, contrary to more recent
observations. A phase shift of the light curve minimum was found
in the infrared also by \citet{Johnson04}.
%
\subsection{Infrared Light Curve}\label{sec:subsec:irLC}
\begin{figure}
\begin{center}
\begin{tabular}{l}\hskip -0.6cm
{\resizebox{9cm}{!}{\includegraphics{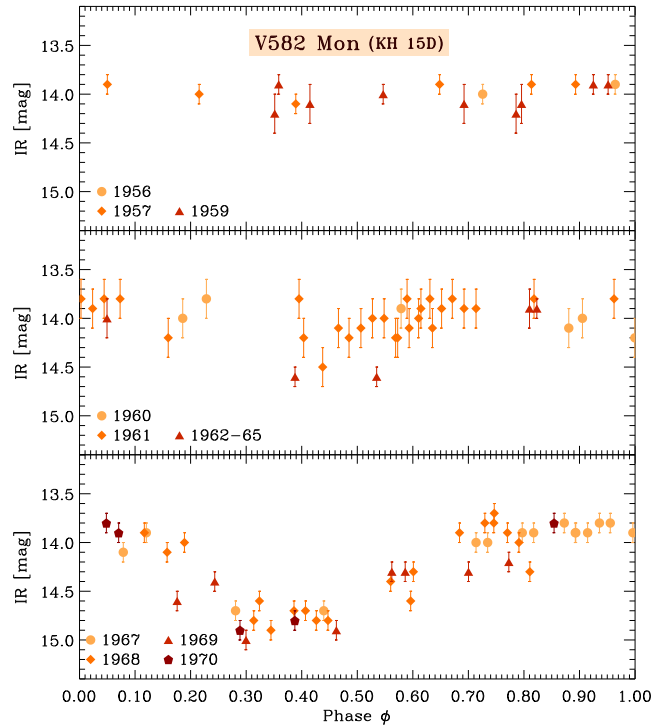}}}\\
\end{tabular}
\end{center}
\vskip -0.3 true cm \caption{Infrared magnitude light curve of
V582 Mon (KH 15D) as a function of the phase $\phi$ from our
1955-1970 data. The spectroscopic binary period (48.38 days) and
the epoch 2452352.26 reported by \citet{Herbst02} are used. All
the data points reported in the plot are real observations,
without any data point duplication. In this plot the eclipses
seems possibly to start at the beginning of 1960s, and the
periodic variability in the subsequent years appear
approximatively sinusoidal, with slow descendent and increasing
stages and no evident plateaus.} \label{fig:phaseplotIR}
\end{figure}

The same behaviour is displayed by the infrared magnitude light
curve in Fig. \ref{fig:phaseplotIR}. In the photographic infrared
(similar to the $I_{c}$ band as mentioned above) after the first
weakening the amplitude increases, beginning in 1968, by 1.2 mag.
The minima are coincident with those observed in the blue. The
phase-constructed light curve shows a flat minimum in the years
1967-1970, similar to that observed in 1995, however it
represents a smaller fraction of the period (0.2) than in the
more recent data (0.4). The observed maximum does not appear as a
pronounced plateau, as seen in recent observations. In addition,
the transition between minimum and maximum values is much slower
(0.2 against 0.07 in phase units); the infrared light curve is
better described by a sinusoidal function rather than by a
step-like structure, as observed in 1996-2002.
\par The infrared maximum appears approximatively at the same
level as in recent years. We do not see a clear weakening, as
instead suggested recently in the infrared magnitudes measured in
the \emph{real} $I$ Cousin band. We must however cautiously
underline that the two photometric systems are not identical. On
the other hand, the variability amplitude increased from 1 mag to
3 mag.
%
\subsection{Colour Indexes}\label{sec:colorindex}
\begin{figure}
\begin{center}
\begin{tabular}{l}\hskip -0.7cm
{\resizebox{9.cm}{!}{\includegraphics{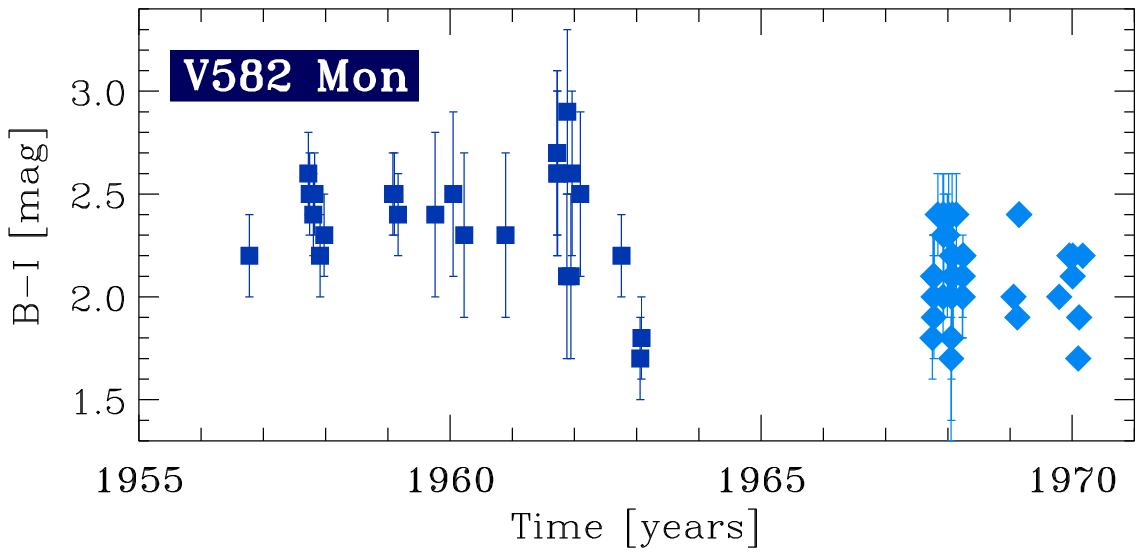}}}\\
\hskip -0.7cm {\resizebox{9.cm}{!}{\includegraphics{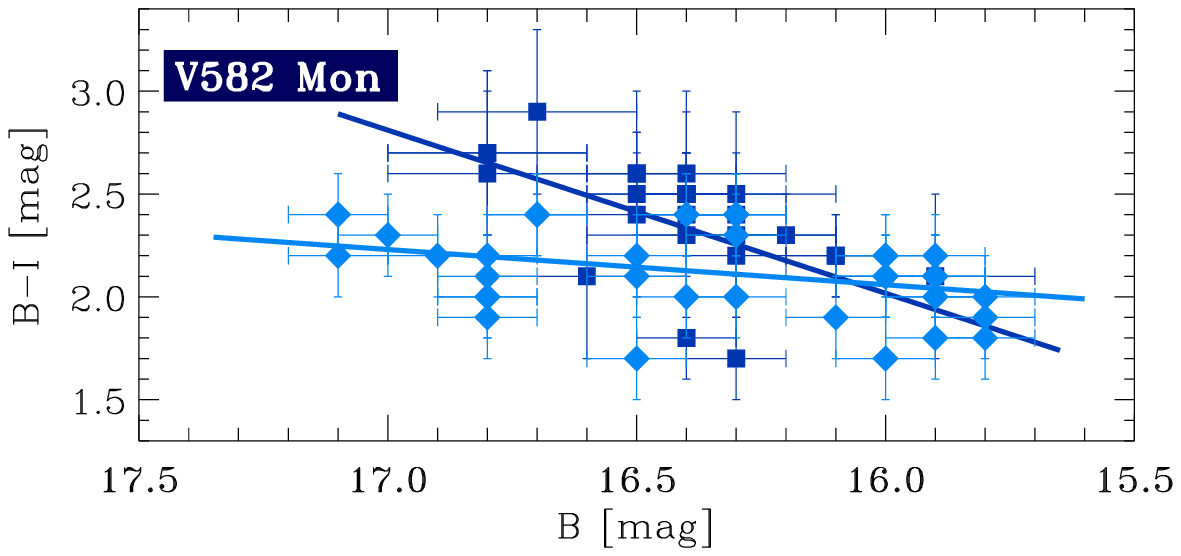}}}\\
\end{tabular}
\end{center}
\vskip -0.3 true cm \caption{Upper panel: the $B-I$ colour index
in function of the time. The plot is clearly in agreement with an
infrared colour excess in the period 1955-1963, implying that the
beginning of the periodical eclipsing variability in the blue was
not followed by the infrared flux that remains rather constant in
those years. Lower panel: the scatter plot between the $B-I$
colour index and the $B$ magnitude brightness shows the same
behaviour. The linear regression trend lines are reported. The
slope is steeper for the 1955-1963 data (blue/darker square
symbols) with respect to the slope of the 1967-1970 data
(azure/lighter diamond symbols) implying that the star becomes
redder when fainter during the 1955-1963 epoch.}
\label{fig:colourindexes}
\end{figure}

The suggested 4-year difference in starting times between the blue
and the IR periodicity is supported by the analysis of the $B-I$
colour indexes. For the calculation of the the $B-I$ index, we
selected only the data from the two bands which were coincident
within an interval of 1 day. Figure \ref{fig:colourindexes} shows
the behaviour of the colour index in time as well as the
correlation with the blue brightness, which fact confirms and
underlines our finding that the beginning of the infrared
variability occurred some years later (possibly 4-years) after
that in the blue. In fact, the $B-I$ index shows a colour excess
and a variability for 1955-1963, while it tends to remain constant
in subsequent years (Fig. \ref{fig:colourindexes} upper panel).
Moreover, the scatter plot $B-I$ versus $B$ shows a steeper trend
for the 1955-1963 data, implying that the star became redder while
fainter during these years. We point out that there are several
parallel B and I observations within one day, to support this
hypotheses. Only in the subsequent years (1967-1970) does the
behaviour of the colour index not depend on the source brightness,
implying that the star shows the same periodical behaviour in the
blue and infrared bands.
%
%
%
%
\section{Discussion and Conclusions}\label{sec:conclusions}
We have presented the results of a long-term series of
observations of the variable star V582 Mon (KH 15D) in the
cluster NGC 2264, performed in the blue and near-infrared bands
between October 13, 1955 and March 1, 1970. This set of
homogeneous data in two colours is unique to our knowledge for
this period and reveals the start of the eclipse phase of V582 Mon
in the two colours.
\par The analysis of our data provides the following important results:
\begin{enumerate}
  \item{} in the blue, the star showed the first
  magnitude variations at the beginning of 1958;
  \item{} in the near infrared, the first magnitude variation was
observed roughly 4 years later;
  \item{} over the above 4 year period the star showed an infrared excess,
  at least around minima.
  \item{} the period did not change;
  \item{} the shape of the light curve appears different from the
  present one, i.e. a sinusoidal function rather than a curve with flat
  minima and maxima, as in \citet{Kearns98}.
\end{enumerate}
A possible model to explain these characteristics and their
evolution would be the following. Until 1958, V582 Mon (KH 15D)
was surrounded by an edge-on disk or gaseous torus, which did not
cause variations in the brightness. Subsequently, thin dust, able
to absorb the blue radiation (but not the IR radiation) began to
form, distributed as a sort of ``banana'' as in the model proposed
by \citet{Barge03}, but with nearby, or extended thinner edges as
indicated by the absence of a flat maximum. Afterwards the dust
aggregated into larger particles, and the selective absorption
became total absorption. Due to the increasing number and
dimension of the dust grains, the observed minima deepen. At the
same time, the ``banana'' dust cloud (or its denser part)
contracted, creating wider, flattened minima and steeper
variations of the brightness to reach the actual shape and values
of the light curve \citep{Winn04}. This suggests the presence of a
body or agglomerate of dust and/or stones less extended but denser
and with sharper edges.
\par These hypotheses imply that in the last 40 years we have
observed the birth of a giant planet near the star, as that
discovered (with other techniques) in other nearby stars
\citep[see, e.g. ][]{Butler03,Marcy98}. More likely, we advance
the notion of the formation of a blob of material, from which one
or more planets could form, or that could dissolve or evolve in an
unforeesable way. This is supported by the fact that V582 Mon (KH
15D) is a pre-main sequence star, i.e. in formation. Our data does
not exclude other models. In any case, any proposed model must
take into account the non-variability prior to 1958, the
progressive increase of the absorption, the interval of 4 years
separating the beginning of the blue and infrared variability and
above all, the transformation from a sinusoidal light curve to an
actual step-like structure characterized by flat maxima and
minima. We believe that the evolution of this unique and peculiar
stellar object must be followed in the future with several
different instruments, and in particular with the same photometric
and spectroscopic techniques used in the past.
%
\section{Acknowledgments}
The authors wish to thank the referee W. Herbst for very useful
comments that helped to improve this work.\\
M. Busso (Perugia University) and W. J. Burger (INFN-Perugia) are
gratefully acknowledged for their contribution in the final
revision of the paper.
%
%
%
%
%
\bibliographystyle{aa}

\label{lastpage}

\end{document}